\documentclass[%
12pt,
 superscriptaddress,
 amsmath,amssymb,
 aps,UTF8,
]{revtex4-1}

\usepackage{graphicx}
\usepackage{dcolumn}
\usepackage{bm}
\usepackage{xcolor}
\usepackage{url}
\usepackage{float}
\usepackage{tabularx}
\usepackage{lipsum}
\usepackage{array}

\usepackage{booktabs}

\usepackage{subfigure}
\usepackage{multirow}

\begin{document}

\title{Production of Tensor Glueball in Ultraperipheral $pp$, $AuAu$ and $PbPb$ Collisions}
\author{Wei Kou}
\email{kouwei@impcas.ac.cn}
\affiliation{Institute of Modern Physics, Chinese Academy of Sciences, Lanzhou 730000, China}
\author{Xurong Chen}
\email{xchen@impcas.ac.cn (Corresponding author)}
\affiliation{Institute of Modern Physics, Chinese Academy of Sciences, Lanzhou 730000, China}
\affiliation{University of Chinese Academy of Sciences, Beijing 100049, China}
\affiliation{Guangdong Provincial Key Laboratory of Nuclear Science, Institute of Quantum Matter, South China Normal University, Guangzhou 510006, China}


\begin{abstract}
	 In this work, we discuss the production of a tensor glueball in the proton-proton and nuclei-nuclei ultraperipheral collisions (UPCs). The cross section is calculated in the equivalent photon approximation (EPA). The total UPCs cross section is related to the matrix elements of $\gamma \gamma\to \pi^0 G_2$ process, which can be described as in terms of neutral pion and tensor glueball $G_2$'s distribution amplitudes. The predictions for the rapidity distributions of UPCs cross sections and total cross sections are presented. The theoretical uncertainty of the whole calculation is considered. We consider the decay modes of the final-state tensor glueball and estimate the corresponding number of events. We estimate that there are about 10 tensor glueball
	 $G_2\to\phi\phi$ signal events in $PbPb$ collisions at the LHC. We also argue that the lead-lead UPC experiments at future HE-LHC and FCC experiments will provide larger opportunities to measure tensor glueballs.
	
\end{abstract}

\pacs{24.85.+p, 13.60.Hb, 13.85.Qk}
\maketitle


\section{Introduction}
\label{sec:intro}
The photon-induced processes at RHIC, Tevatron and LHC have motivated a series of studies, which test the validity of the Standard Model and deepen our understanding of the Beyond Standard Model (BSM) physics. The analysis of different final
states produced in $\gamma\gamma$ and $\gamma h$ interactions at hadronic collisions are fully discussed e.g. in the review \cite{LHCForwardPhysicsWorkingGroup:2016ote}. The hadronic states made up only from gluons are known as glueballs \cite{Fritzsch:1972jv,Fritzsch:1973pi,Fritzsch:1975tx}. The first study to give the glueball mass came from bag model \cite{DeGrand:1975cf}. The spectrum of glueball states have been determined by lattice QCD calculations, see Refs. \cite{Bali:1993fb,Morningstar:1999rf,Chen:2005mg,Richards:2010ck,Gregory:2012hu,Athenodorou:2020ani}. Because of the special nature of glueballs under SU(3) symmetry, which have the same quantum number as a meson state composed of quark-antiquark, finding glueball in experiments is a challenge. The analytical approximation of QCD predicts a scalar glueball at 1850 $\sim$ 1980 MeV \cite{Szczepaniak:2003mr,Huber:2020ngt,Rinaldi:2021dxh}. Unlike scalar glueballs, which have some scalar candidates \cite{Klempt:2007cp,Crede:2008vw,Ochs:2013gi}, much less is even known about tensor glueballs. The mass spectra of the lowest $2^{++}$ tensor glueballs were analyzed from the lattice QCD, and these works give the masses around 1900 $\sim$ 2600 MeV \cite{Klempt:2022vxs}. We note there are some experimental evidences that such state have been seen in various processes \cite{Etkin:1985se,Belle:2013eck,BESIII:2016qzq,Klempt:2022vxs}. In particular, the recent work has shown the related analysis of scalar glueball as well as tensor glueball \cite{Klempt:2022vxs}. 

In hadron-hadron collisions, ultraperipheral collisions (UPCs) are important in the theoretical and experimental investigations \cite{Baur:2001jj,Bertulani:2005ru,Baltz:2007kq}. In hadron–hadron
UPCs, the impact parameter between the two hadrons is larger than the sum of
the two hadron's radius. Consequently, strong interaction between the hadrons is
suppressed for the large distance. On the other hand, photon can be emitted from
the hadrons at high energy, thus, the photon can interact with the hadron or the
photon emitted from the other hadron. There has been a lot of published works on the topic of glueball production by UPCs process \cite{Schramm:1999tt,Machado:2010qg,Klusek-Gawenda:2013rtu}.
These studies include the calculation of cross sections for scalar and tensor glueballs, and some of these studies have estimated the production of glueballs in peripheral collisions by strong double diffractive scattering, i.e. Pomeron-Pomeron (PP) exchange \cite{Schramm:1999tt,Machado:2010qg}. In the present work,  we consider the glueball production in two-photon fusion which can be calculated using the narrow-resonance approximation \cite{Schramm:1999tt,Machado:2010qg}. 

The production cross section in our method is computed using
the QCD factorization approach \cite{Lepage:1980fj,Brodsky:1981rp,Efremov:1979qk,Kivel:2017tgt}. In Ref. \cite{Kivel:2017tgt}, the authors studied the production of
tensor $2^{++}$ glueballs in two-photon collisions at high momentum
transfer. One considered that the coupling of quarks
and gluons to the final mesonic states is described by the distribution amplitudes (DAs) describing the momentum fraction distribution of partons at zero transverse separation in a two-particle Fock state \cite{Kivel:2017tgt}. We review the details of this part in the following section.

Based on the above discussion, tensor glueballs can be considered to be produced during the UPCs process and a significant number of events may be observed in future heavy ions collision experiments.
The calculation of $\gamma\gamma\to \pi^0 G_2$ process and the corresponding UPCs process are reviewed in Sec. \ref{sec:form}. The results for the total cross sections and rapidity distributions are shown in Sec. \ref{sec:dissc}. Summary and outlook are presented in the final section.

\section{$\gamma\gamma\to \pi^0 G_2$ to Ultraperipheral Collisions}
\label{sec:form}
\begin{figure}[htbp]
	\centering  
	\subfigure{
		\label{fig:gluon}
		\includegraphics[width=0.4\textwidth]{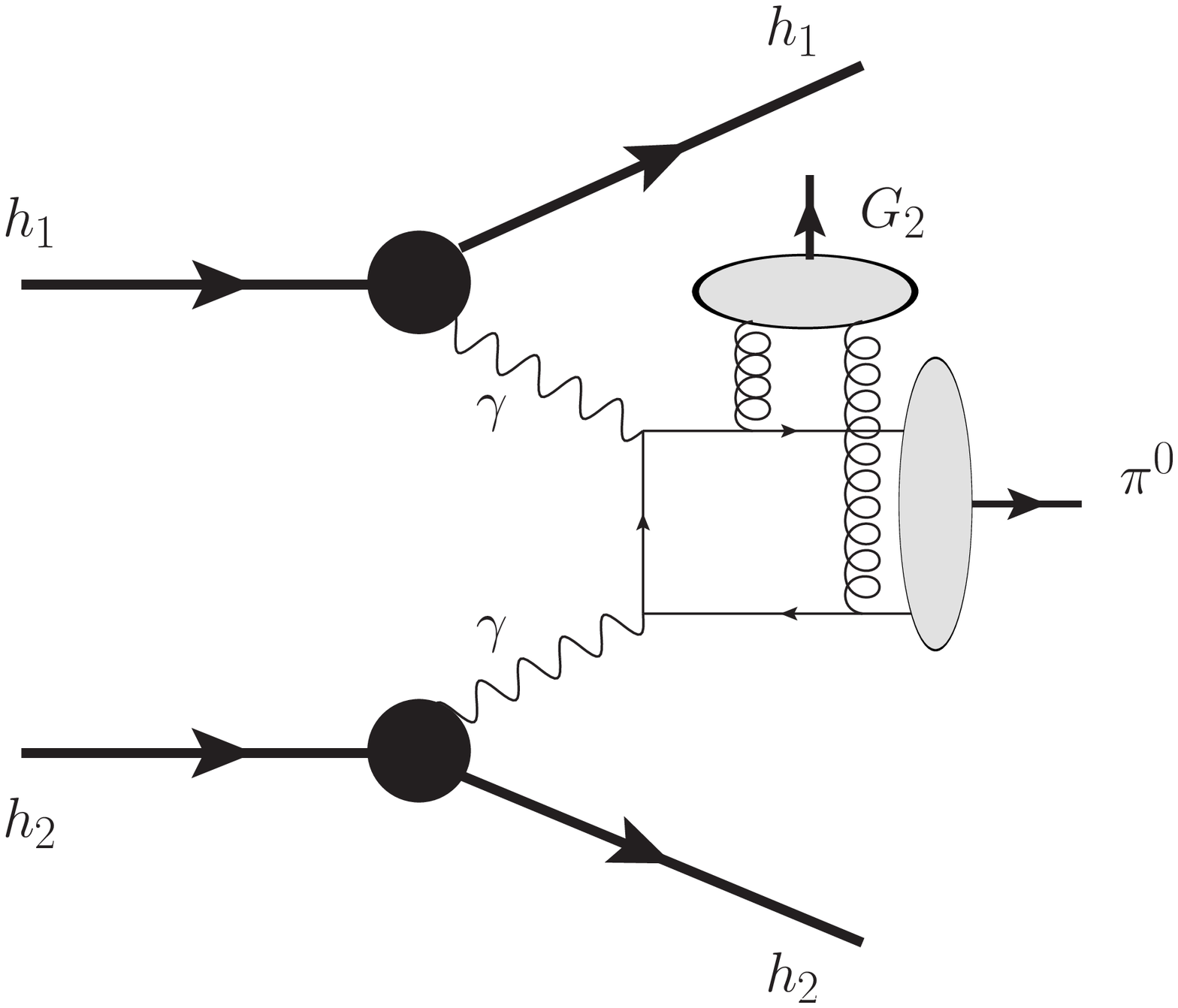}}
	\subfigure{
		\label{fig:quark}
		\includegraphics[width=0.4\textwidth]{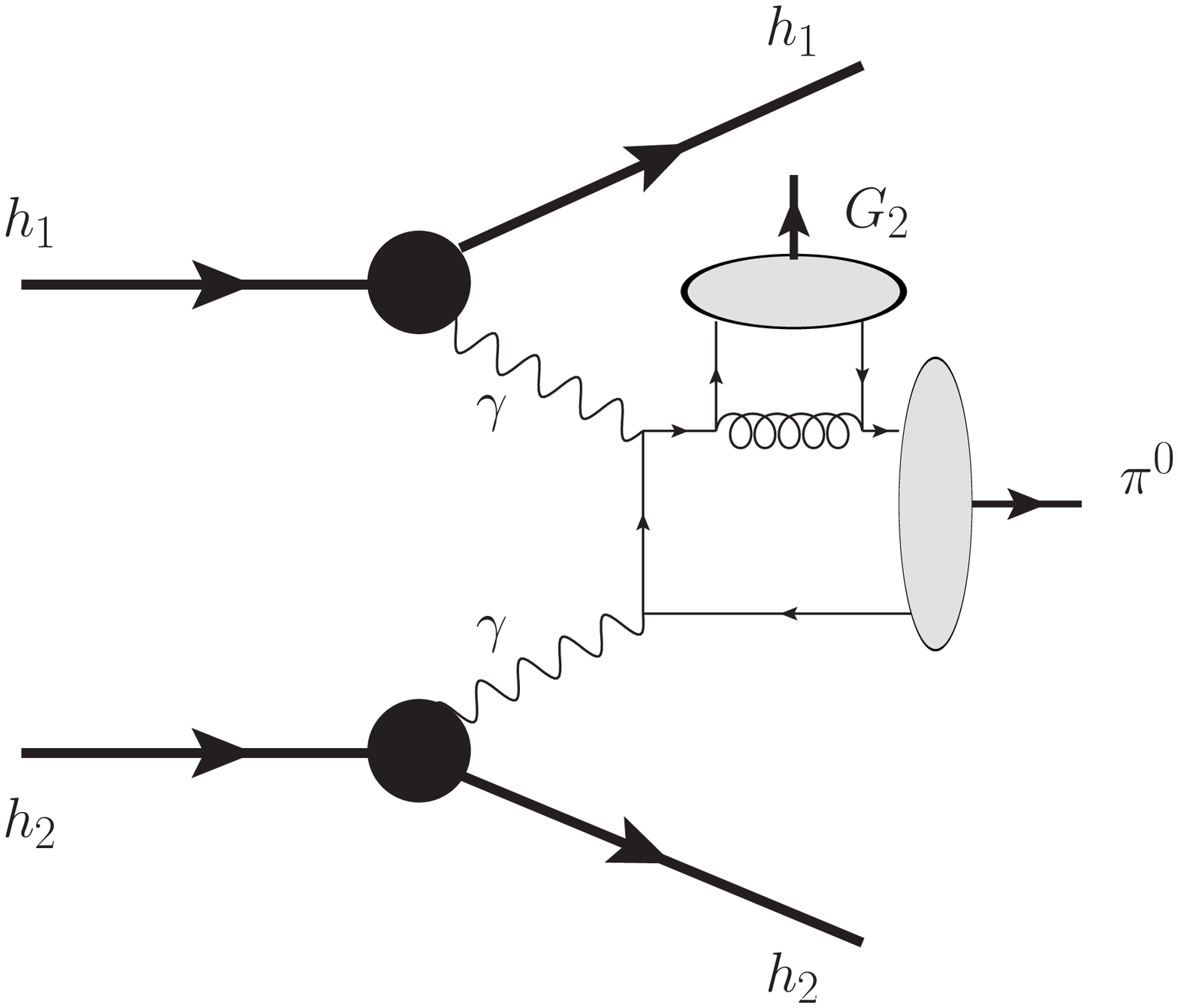}}
	\caption{Two typical diagrams of $h_1h_2\to h_1h_2\pi^0G_2$. The blobs denote the
		distribution amplitudes discussed in Ref. \cite{Kivel:2017tgt}. $h_1,\ h_2$ represent protons or nucleus.}
	\label{fig:Feynman}
\end{figure}

As a starting point for the core, one can recall the two-photon collisions in Ref. \cite{Kivel:2017tgt} that produce $\pi^0$ and tensor glueball $G_2$. The production amplitude for the $\gamma\gamma\to \pi^0 G_2$ process should be described in terms of the helicity amplitudes \cite{Kivel:2017tgt}
\begin{equation}
	\begin{aligned}
		&i A_{\pm \pm}= \varepsilon_{1 \mu}(\pm) \varepsilon_{2 \nu}(\pm) \\
		& \times \int d^{4} x e^{-i\left(q_{1} x\right)}\left\langle G_{2}(p), \pi^{0}(k)\left|T\left\{J_{\mathrm{em}}^{\mu}(x), J_{\mathrm{em}}^{\nu}(0)\right\}\right| 0\right\rangle,
	\end{aligned}
	\label{eq:amplitude}
\end{equation}
where $p$ and $k$ denote the momenta of glueball $G_2$ and $\pi^0$. $\varepsilon_{1 \mu}$ and $\varepsilon_{2 \nu}$ are the polarization vectors of initial photons. The electromagnetic currents are described as $J_{\mathrm{em}}^{\mu}(x)$ and $J_{\mathrm{em}}^{\nu}(0)$ (see FIG. \ref{fig:Feynman}).
From the helicity amplitudes, the differential cross section is written as \cite{Budnev:1975poe}
\begin{equation}
	\frac{d \sigma_{\gamma \gamma}\left[\pi^{0} G_{2}\right]}{d \cos \theta}=\frac{1}{64 \pi} \frac{W_{\gamma\gamma}+m^{2}}{W_{\gamma\gamma}^{2}}\left(\left|\overline{A_{++}}\right|^{2}+\left|\overline{A_{+-}}\right|^{2}\right),
	\label{eq:diff-xsec}
\end{equation}
where $\theta$ is the scattering angle in the c.m.s., $W_{\gamma\gamma}$ denotes the $\gamma\gamma$ c.m.s. energy and one can set $|A_{++}|=|A_{--}|$ and, $|A_{-+}|=|A_{+-}|$ \cite{Kivel:2017tgt}. $m$ is the mass of tensor glueball and we choose $m=2.3$ GeV ($f_2(2300)$ or $f_2(2340)$) which have been recently observed by the Belle \cite{Belle:2013eck} and BESIII \cite{BESIII:2016qzq} collaborations are good candidates to be tensor glueball. The sum over polarization of amplitudes are
\begin{equation}
	\left|\overline{A_{+\pm}}\right|^{2}=\sum_{\lambda=-2}^{2} A_{+\pm}(\lambda) A_{+\pm}^{*}(\lambda),
	\label{eq:polar}
\end{equation}
where $\lambda$ is the polarization of the glueball.
The authors in Ref. \cite{Kivel:2017tgt} also introduced light-like vectors and some approximations. They considered two diagrams (see the internal parts in FIG. \ref{fig:Feynman}). The blobs in FIG. \ref{fig:Feynman} denote the light-cone matrix elements which define the DAs of the outgoing meson and glueball. One can finally get the polarized amplitudes when the colliding photons have the same helicities. It can be treated with the DAs (see eq. (15-19) in Ref. \cite{Kivel:2017tgt} for details). The total cross section of two-photons interactions is $\sigma_{\gamma \gamma}(W_{\gamma\gamma})=\int d\cos\theta \frac{d\sigma_{\gamma \gamma}(W_{\gamma\gamma})}{d \cos\theta}$. 

 Photon interactions lead to a wide variety of final states. They couple to all charged particles, including leptons, quarks, and charged gauge bosons \cite{Hencken:1995me,Baur:2001jj}. The production of massive tensor glueball $G_2$ in hadronic collisions is represented in FIG. \ref{fig:Feynman}. Both incident hadrons can be considered as the source of photons. 

In UPCs, the total cross section of $h_1h_2\to h_1h_2\pi^0G_2$ can be written in the well-known form form \cite{Baur:1990fx}
\begin{equation}
	\begin{aligned}
		&\sigma (h_1h_2 \rightarrow h_1 h_2 \pi^0G_2  ; s_{h_1 h_2}) \\
		&= \int d^{2} \mathbf{b}_{1} d^{2} \mathbf{b}_{2} \theta(|\boldsymbol{b}_{1}-\mathbf{b}_{2}|-(R_{h_1}+R_{h_2})) \frac{d \omega_{1}}{\omega_{1}} \frac{d \omega_{2}}{\omega_{2}} \\
		&\times N_{h_1}\left(\omega_{1}, \boldsymbol{b}_{1}\right) N_{h_2}\left(\omega_{2}, \mathbf{b}_{2}\right) \hat{\sigma}(\gamma \gamma \rightarrow \pi^{0} G_2; 4 \omega_{1} \omega_{2}),
		\label{eq:total xsec}
	\end{aligned}
\end{equation}
where $\boldsymbol{b}_{1}$ and $\boldsymbol{b}_{2}$ represent the impact parameters, $R_{h_{1(2)}}$ denotes the hadron or nucleus' radius. For proton, the radius $R_p=0.84$ fm, for nucleus we choose $R_A = r_0 A^{\frac{1}{3}}$ with $r_0 = 1.2$ fm. In particular, $N(\omega,\boldsymbol{b})$ is the equivalent photon flux for a
given photon energy $\omega$ and impact parameter $\boldsymbol{b}$, which can
be expressed in terms of the form factor $F(q^2)$ for the equivalent photon source as follows
\begin{equation}
	\begin{aligned}
		&N_{A}(\omega, b)=\frac{Z^{2} \alpha_{e m}}{\pi^{2}} \frac{1}{b^{2} \omega} \\
		&{\left[\int u^{2} J_{1}(u) F\left(\sqrt{\frac{(b \omega / \gamma_L)^{2}+u^{2}}{b^{2}}}\right) \frac{1}{(b \omega / \gamma_L)^{2}+u^{2}} \mathrm{~d} u\right]^{2} },
	\end{aligned}
\label{eq:flux}
\end{equation}
where $Z$ is the proton number of nucleus and $\alpha_{e m}$ denotes the fine structure constant. $\gamma_L$ is the Lorentz factor which is discussed below. $J_n(u)$ is the first kind Bessel function. Form factor is the Fourier transform of charge distribution in the nucleus. If one assume $\rho(r)$  is the spherical symmetric charge distribution, the form factor is a function
of photon virtuality $q^2$ \cite{Klusek-Gawenda:2010vqb}.

\begin{equation}
	F\left(q^2\right)=\int \frac{4 \pi}{q} \rho(r) \sin (q r) r d r=1-\frac{q^2\left\langle r^2\right\rangle}{3 !}+\frac{q^4\left\langle r^4\right\rangle}{5 !} \cdots.
	\label{eq:rr}
\end{equation}
When the form factor $F(q^2)=1$, it means that the charge distribution is point-like, the integration can be computed analytically. After analytical integration, the equivalent photon flux is given as follows \cite{Bertulani:2005ru}
\begin{equation}
	N(\omega, \boldsymbol{b})=\frac{Z^2 \alpha_{\mathrm{em}}}{\pi^2} \frac{\omega}{\gamma_L^2} K_1^2(|b| \omega / \gamma_L).
	\label{eq:mod1}
\end{equation}
In the above formula, $K_1(x)$ is the second Bessel function. In this paper, we denote
the above equivalent photon flux as model-1.

Moreover, if the charge distribution is not point-like, the form factor can be
expressed in other functions. We refer Refs. \cite{Klusek-Gawenda:2010vqb,Hencken:1995me} then the form factor function is given as 
\begin{equation}
	F(q)=\frac{\Lambda^{2}}{\Lambda^{2}+q^{2}}
	\label{eq:ffs}
\end{equation}
with $\Lambda=0.088$ GeV for nucleus \cite{Goncalves:2012cy,Goncalves:2015hra,Goncalves:2018yxc} and $\Lambda=0.71$ GeV for protons \cite{Xie:2018ict}. If the form factor function is taken as above formula, the equivalent photon flux is
written as
\begin{equation}
	N(\omega, \boldsymbol{b})=\frac{Z^2 \alpha_{\mathrm{em}}}{\pi^2} \frac{1}{\omega}\left[\frac{\omega}{\gamma_L} K_1\left(b \frac{\omega}{\gamma_L}\right)-\sqrt{\frac{\omega^2}{\gamma_L^2}+\Lambda^2} K_1\left(b \sqrt{\frac{\omega^2}{\gamma_L^2}+\Lambda^2}\right)\right]^2 .
	\label{eq:mod2}
\end{equation}
where $K_1$ is the modified Bessel function of the second kind. In this paper, we denote
the above equivalent photon flux as model-2.
 Lorentz factor $\gamma_L$ is computed as $\gamma_L=\sqrt{s_{NN}}/2m_N$, $m_N$ is the nucleon mass and the rapidity-type variable $Y$ defined through
\begin{equation}
	\begin{aligned}
		&\omega_{1}=\frac{W_{\gamma \gamma}}{2} e^{Y}, \quad \omega_{2}=\frac{W_{\gamma \gamma}}{2} e^{-Y}, \\
		&\frac{d \omega_{1}}{\omega_{1}} \frac{d \omega_{2}}{\omega_{2}}=2 \frac{d W_{\gamma \gamma}}{W_{\gamma \gamma}} d Y .
	\end{aligned}
\end{equation}

In next section, we consider the $pp$ and $PbPb$ collisions at LHC and $AuAu$ collisions at RHIC. We also predict the planned c.m.s. energies for the next run of the LHC, as well for the future High-Energy LHC \cite{FCC:2018bvk} and Future Circular Collider (FCC) \cite{FCC:2018vvp}.

\section{Results and discussion }
\label{sec:dissc}

\begin{figure}[htbp]
	\centering  
	\includegraphics[width=0.6\textwidth]{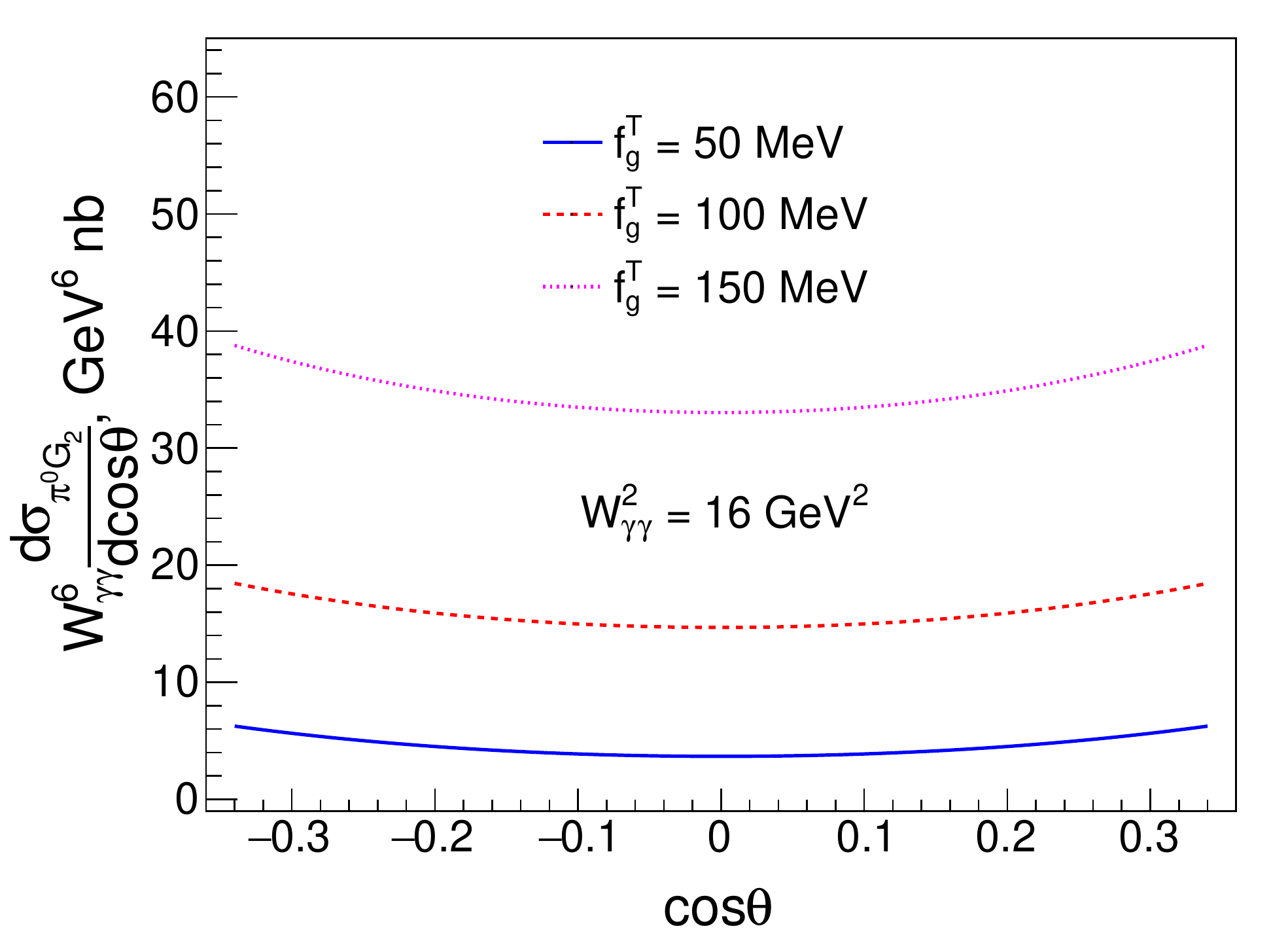}
	\caption{The example of cross section as a function of $\cos\theta$ at $W_{\gamma\gamma}^2=16$ GeV$^2$. The solid, dashed and dotted lines correspond to $f_g^T=50$ MeV, 100 MeV and 150 MeV \cite{Kivel:2017tgt}.}
	\label{fig:dsigma-2gamma}
\end{figure}

We first discuss the scattering matrix element of the $\gamma\gamma\to G_2\pi^0$ process and the corresponding differential cross section calculation. In Ref. \cite{Kivel:2017tgt} the authors considered the mesonic DAs of pion and tensor glueball. The DA of $2^{++}$ glueball is treated as the tensor meson, see, e.g. Ref. \cite{Braun:2015axa}. In general case there are three light-cone matrix elements which
define two gluon DAs and one quark DA \cite{Kivel:2017tgt}. Constructing the formalization of the glueball DA is more complicated, and it is not the focus of this work. We give here only the simple conclusion in Ref. \cite{Kivel:2017tgt} -- the DAs' model of the glueball depends formally on several coupling constants, $f_g^S$, $f_g^T$ and $f_q$. It is natural to assume that the glueball state strongly overlaps with the gluon wave function and the value of the gluon couplings are
relatively large and can be of the same order as the quark coupling $f_q\sim100$ MeV for quark-antiquark mesons, i.e. $f_g^S\sim f_g^T\sim100$ MeV. After our tests, the helicity amplitude $|A^{+-}|$ containing the coupling constants $f_q$ and $f_g^S$ is always about two orders of magnitude smaller than the amplitude $|A^{++}|$ containing $f_g^T$. One is therefore able to conclude that the contribution with $|A^{+-}|$ does not provide significant numerical impact \cite{Kivel:2017tgt}. Hence the cross section (\ref{eq:diff-xsec}) is only sensitive to the value of tensor coupling $f_g^T$. This can also be seen from the analysis of the $G_2\to \phi\phi$ decay, which can be used for the identification of the glueball state \cite{BESIII:2016qzq}. Based on the above discussion, we give an example of the calculation of the differential cross section (\ref{eq:diff-xsec}). In FIG. \ref{fig:dsigma-2gamma}, we take the coupling $f_g^T=50$, 100, 150 MeV, respectively and set $W_{\gamma\gamma}^2=16$ GeV$^2$. The other parameters that appear in the helicity amplitude we are consistent with Ref. \cite{Kivel:2017tgt}.

\begin{figure}[H]
	\centering  
	\subfigure{
		\label{fig:au-mod1}
		\includegraphics[width=0.47\textwidth]{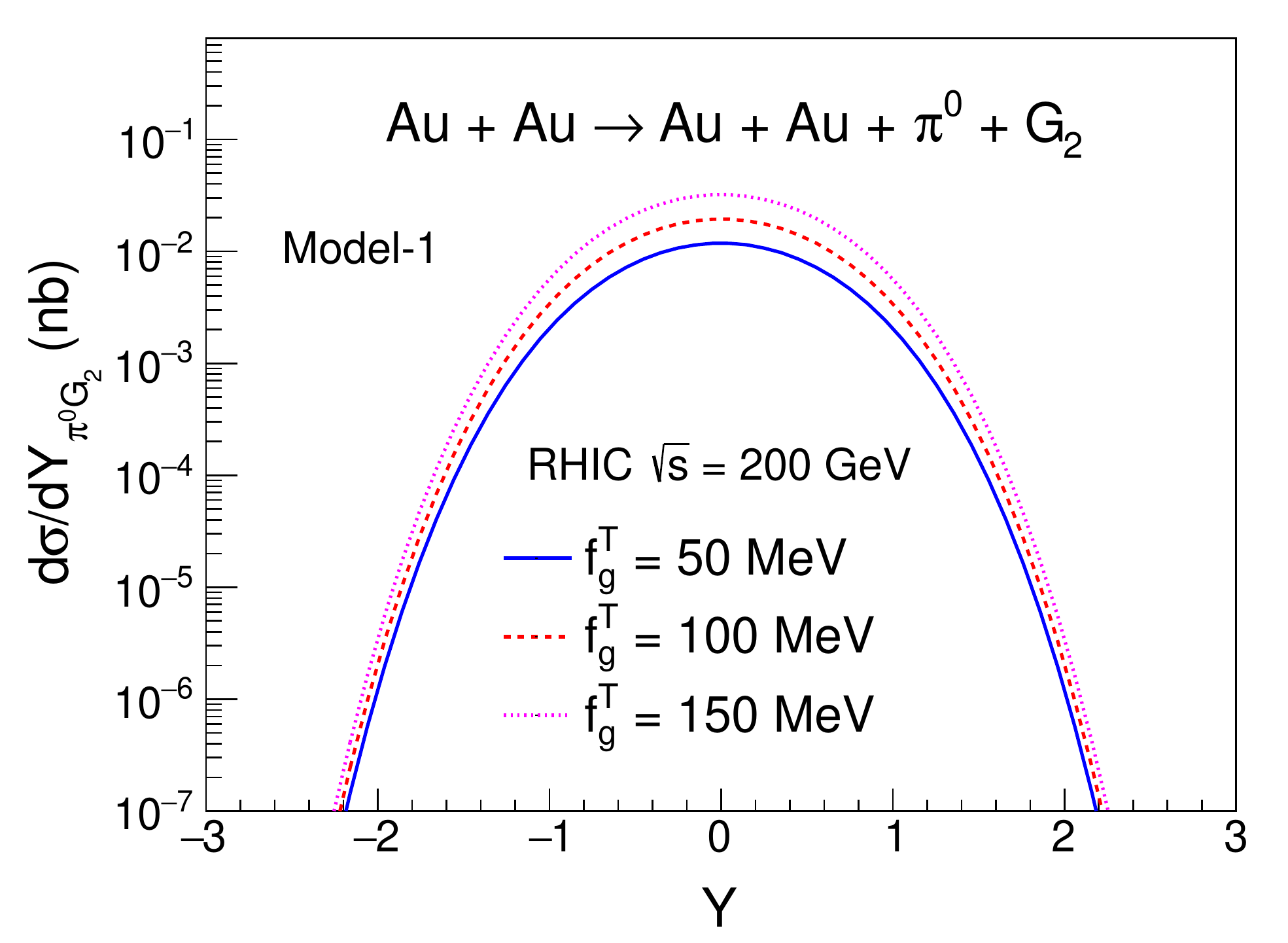}}
	\subfigure{
		\label{fig:au-mod2}
		\includegraphics[width=0.47\textwidth]{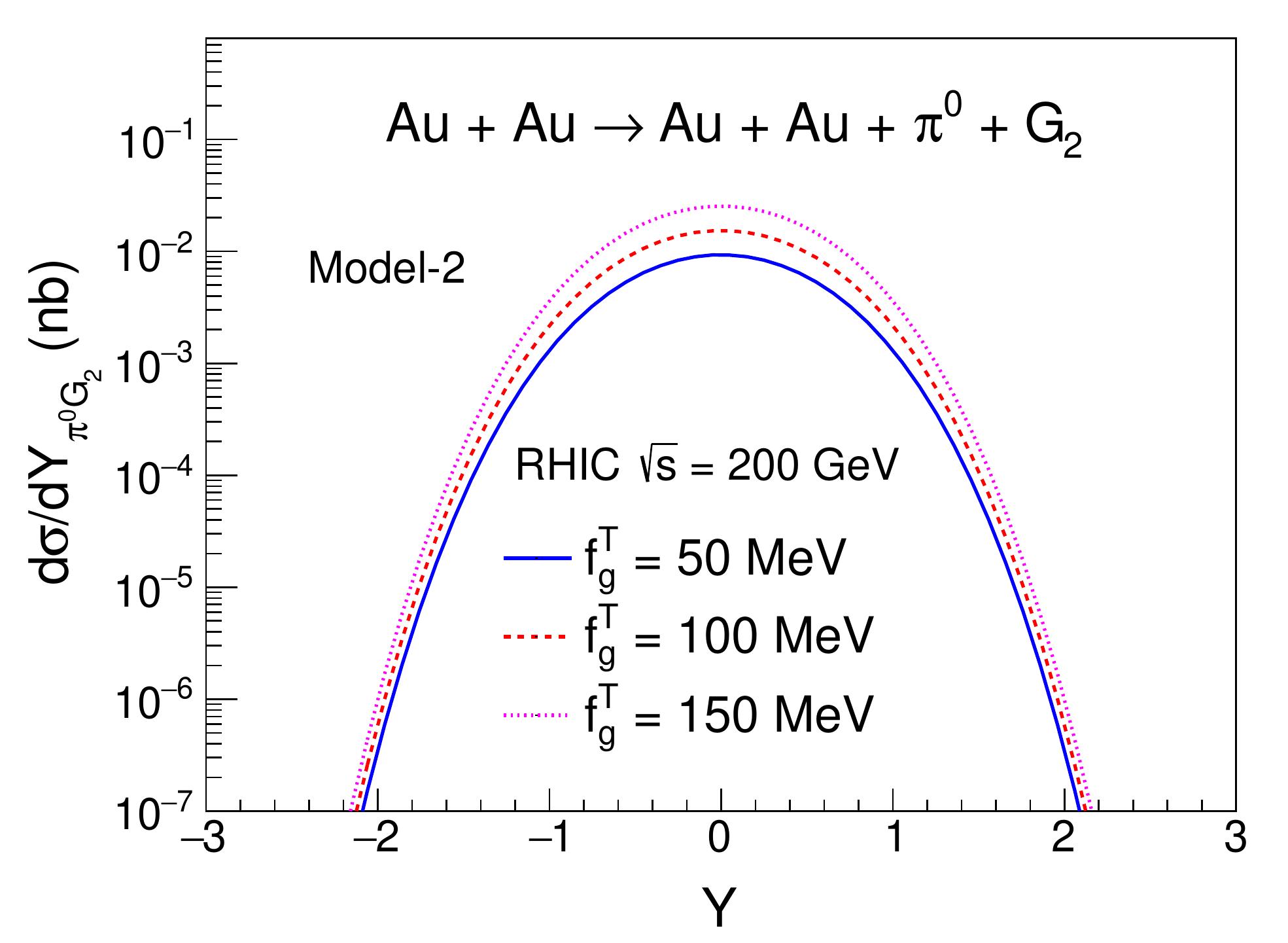}}
	\caption{Rapidity distributions for the UPC process $Au+Au\to Au+Au+G_2+\pi^0$ at RHIC with model-1 (left panels) and model-2 (right panels). The center of mass energies is $\sqrt{s}=200$ GeV. The solid, dashed and dotted lines correspond to $f_g^T=50$ MeV, 100 MeV and 150 MeV \cite{Kivel:2017tgt}.}
	\label{fig:dsigmadY-au}
\end{figure}

\begin{figure}[H]
	\centering  
	\subfigure{
		\label{fig:p1-mod1}
		\includegraphics[width=0.47\textwidth]{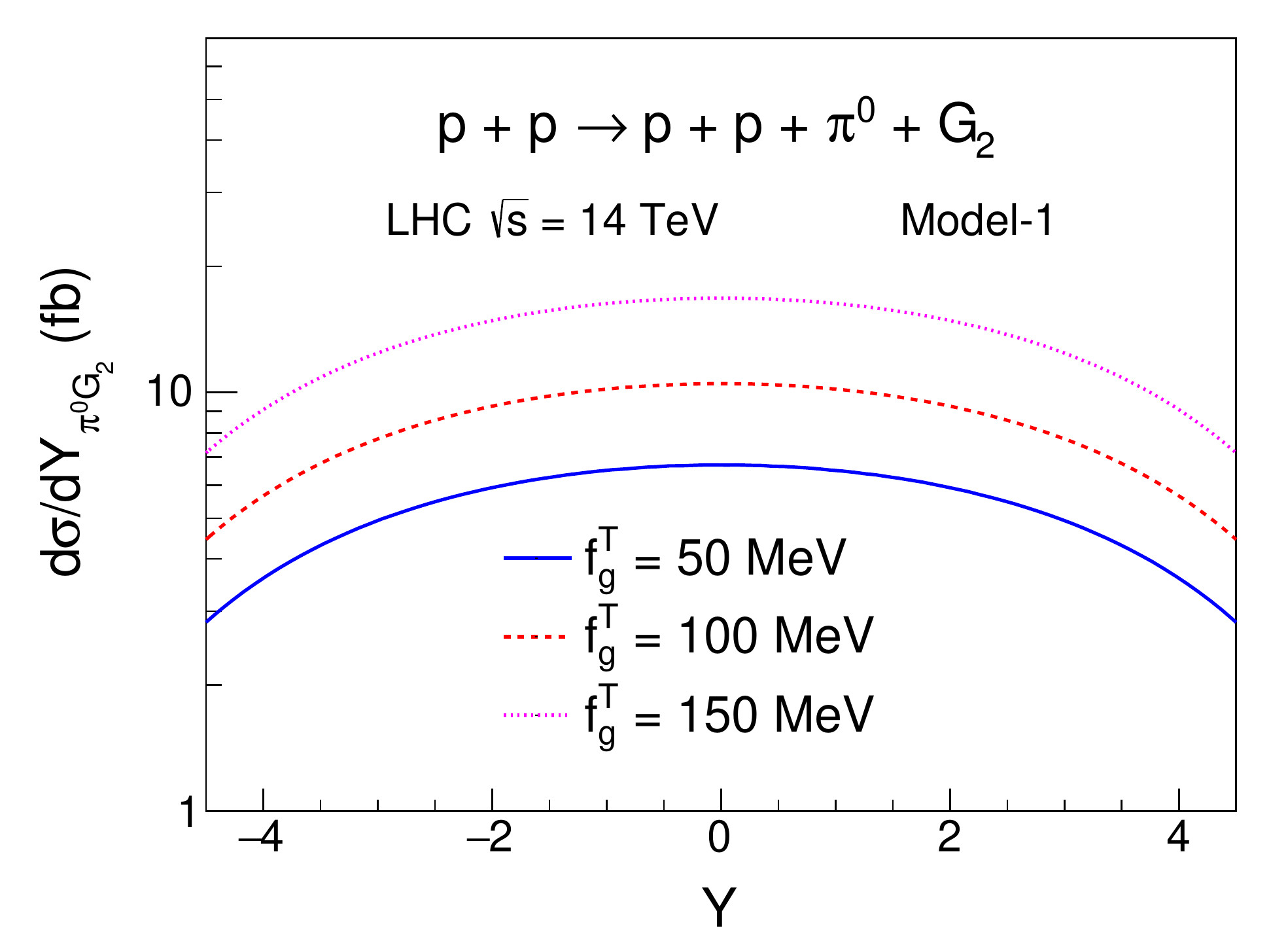}}
	\subfigure{
		\label{fig:p1-mod2}
		\includegraphics[width=0.47\textwidth]{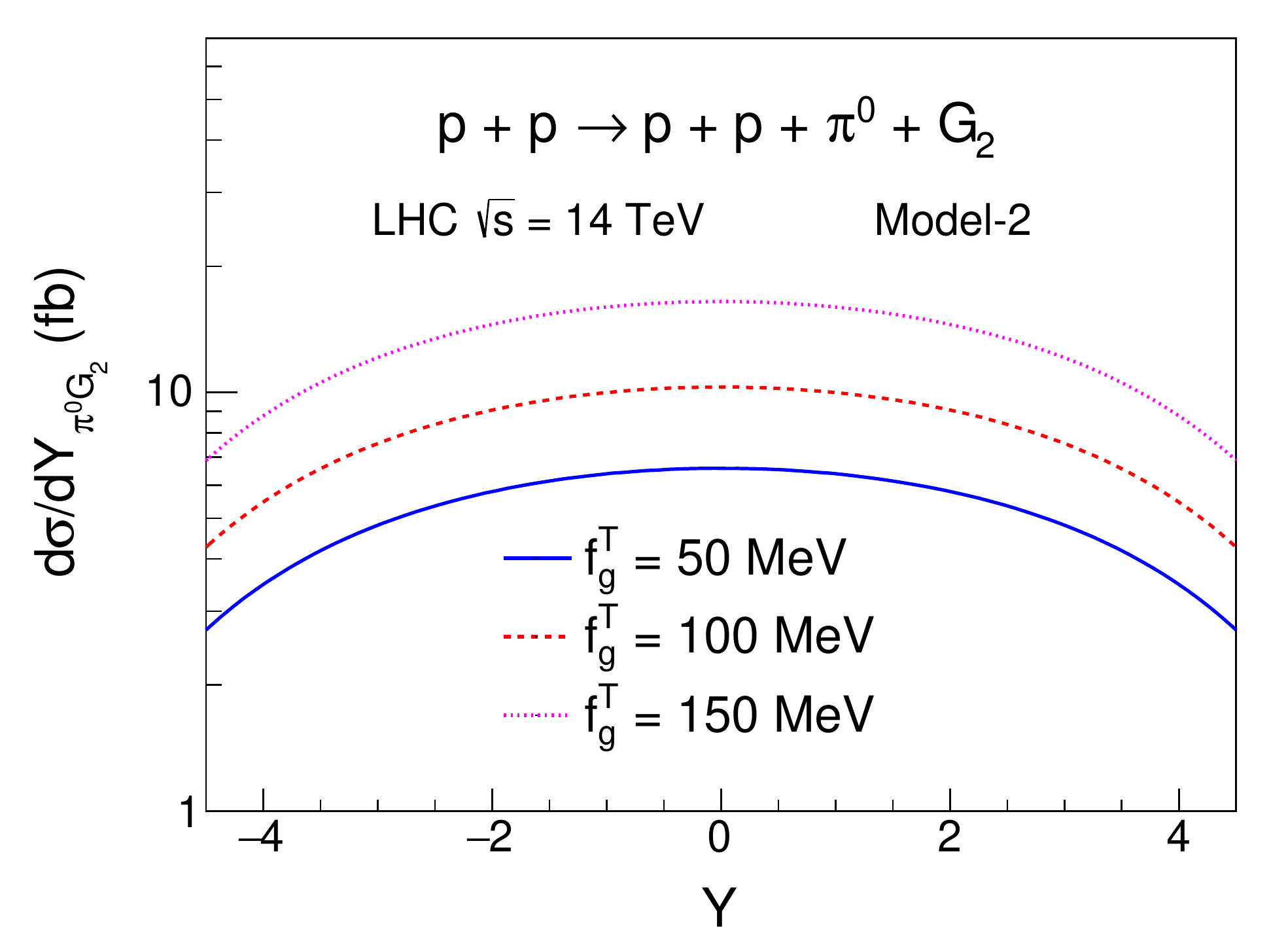}}
	\subfigure{
		\label{fig:p2-mod1}
		\includegraphics[width=0.47\textwidth]{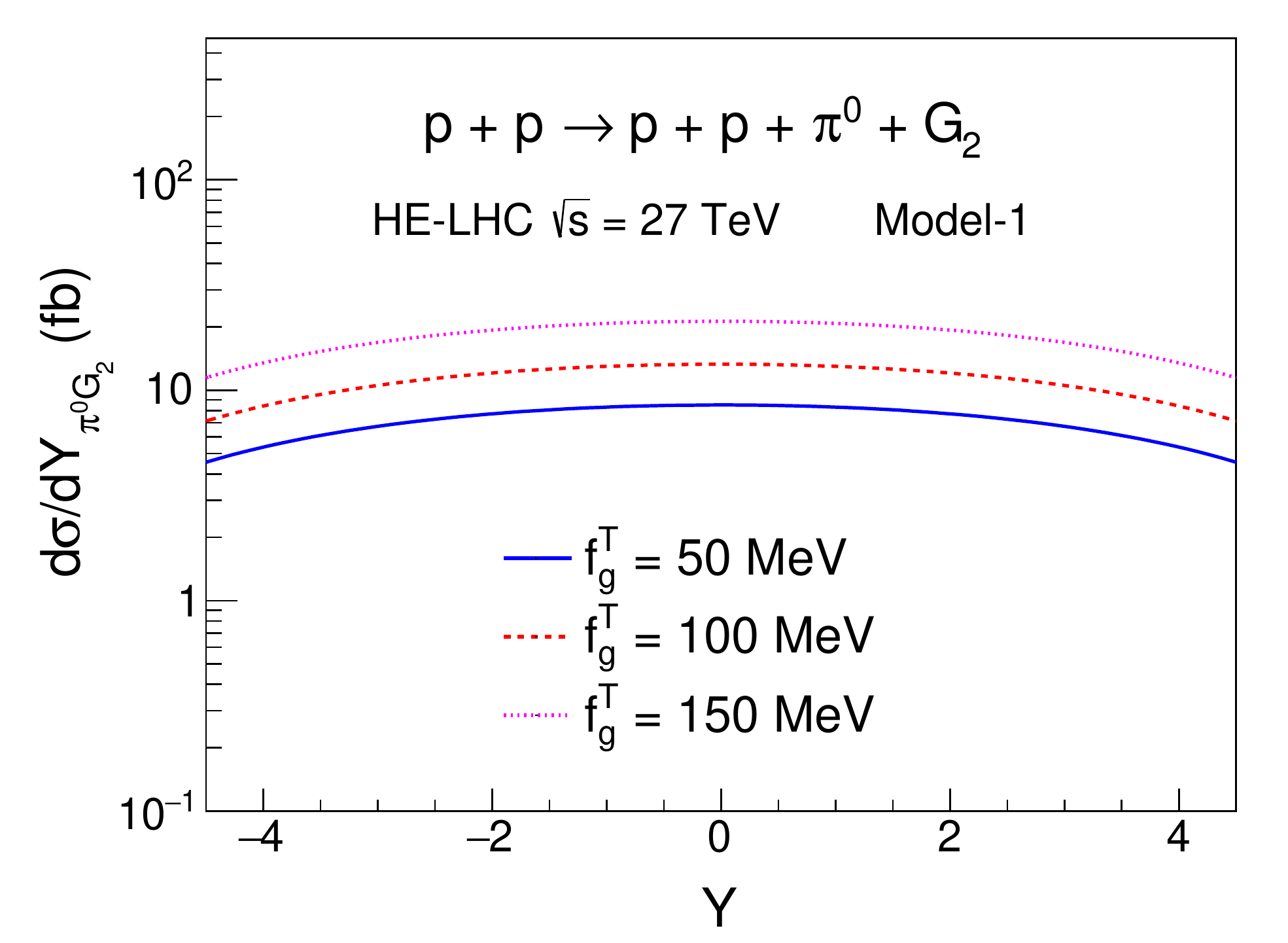}}
	\subfigure{
		\label{fig:p2-mod2}
		\includegraphics[width=0.47\textwidth]{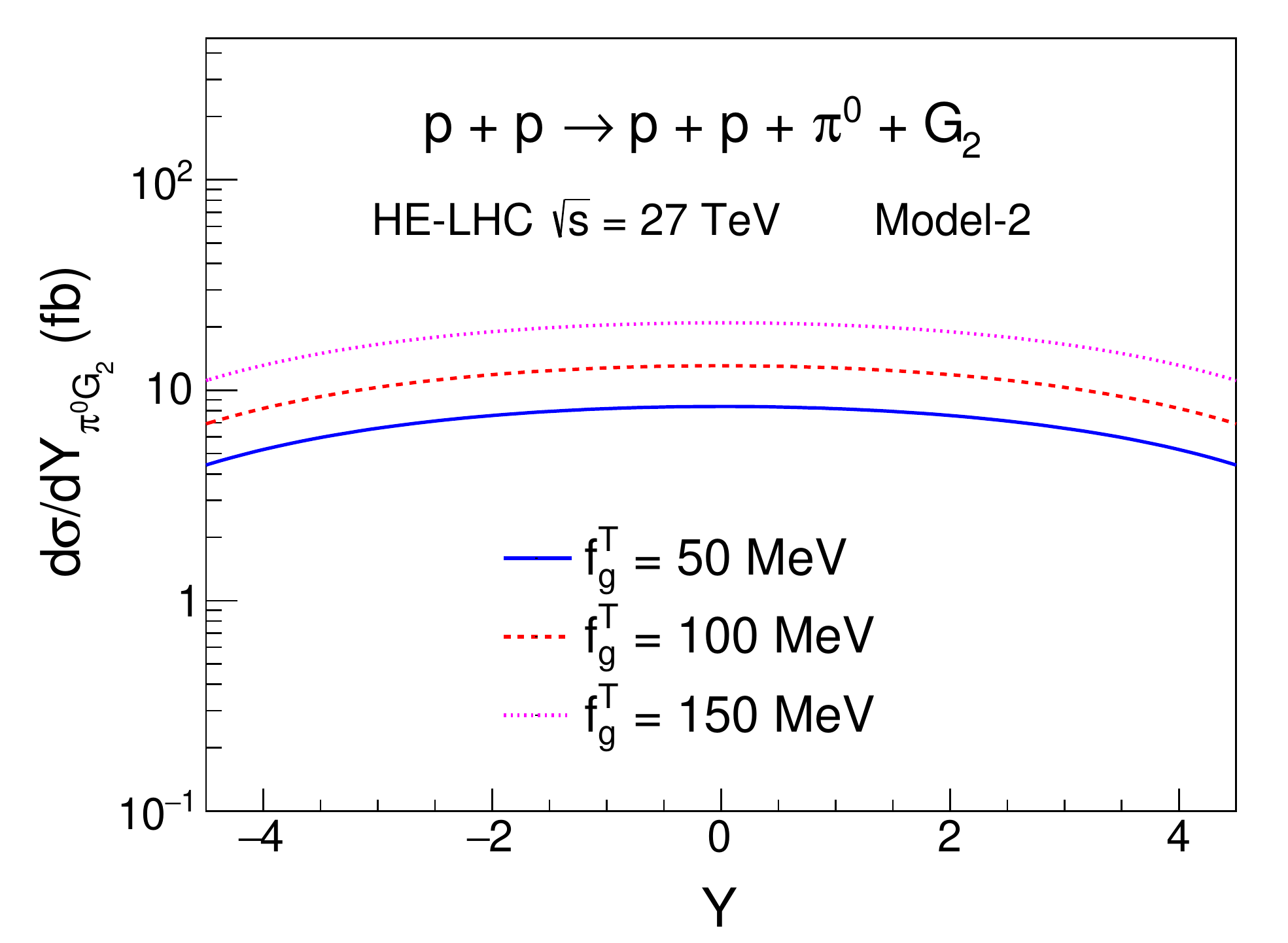}}
	\subfigure{
		\label{fig:p3-mod1}
		\includegraphics[width=0.47\textwidth]{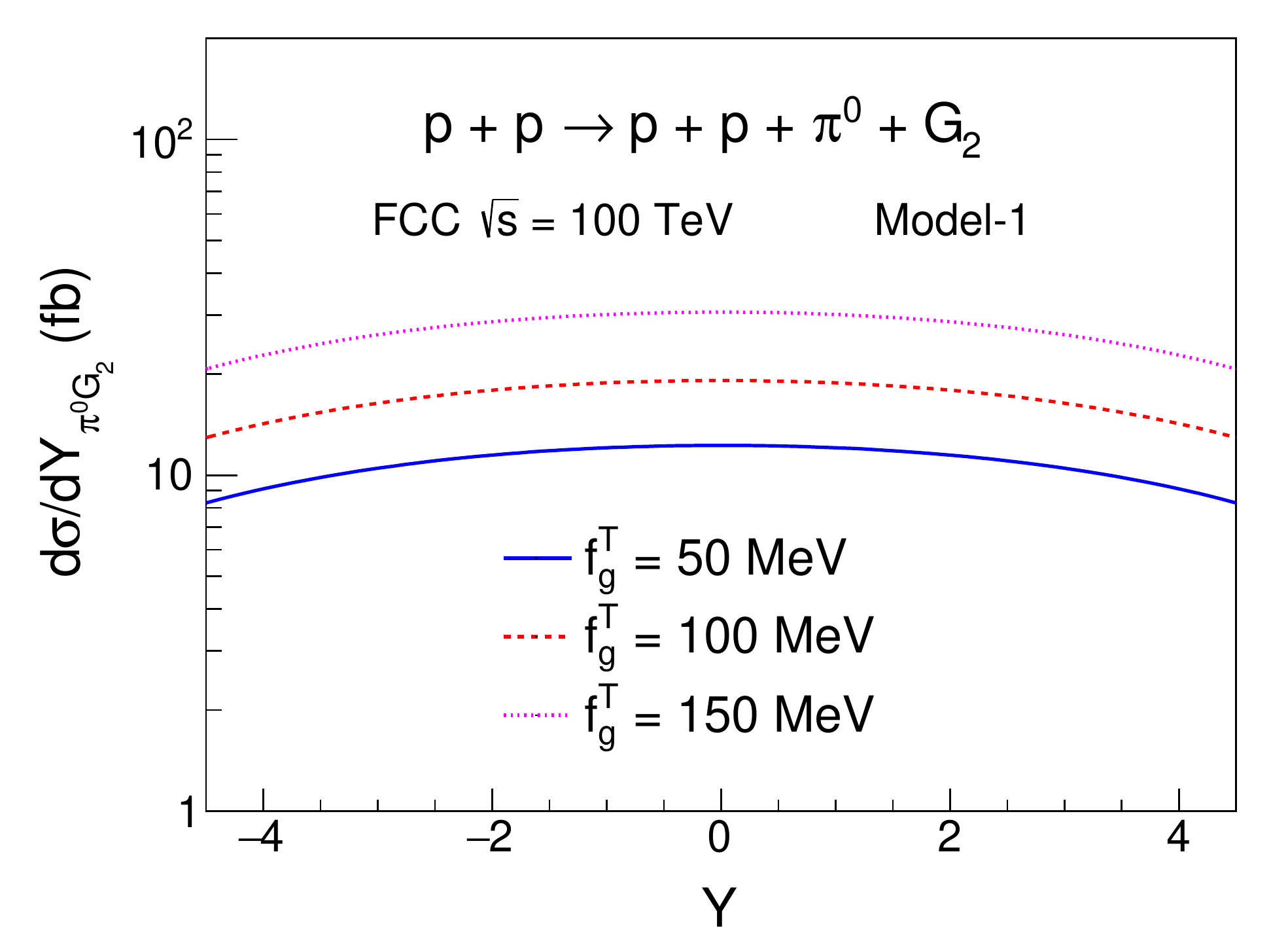}}
	\subfigure{
		\label{fig:p3-mod2}
		\includegraphics[width=0.47\textwidth]{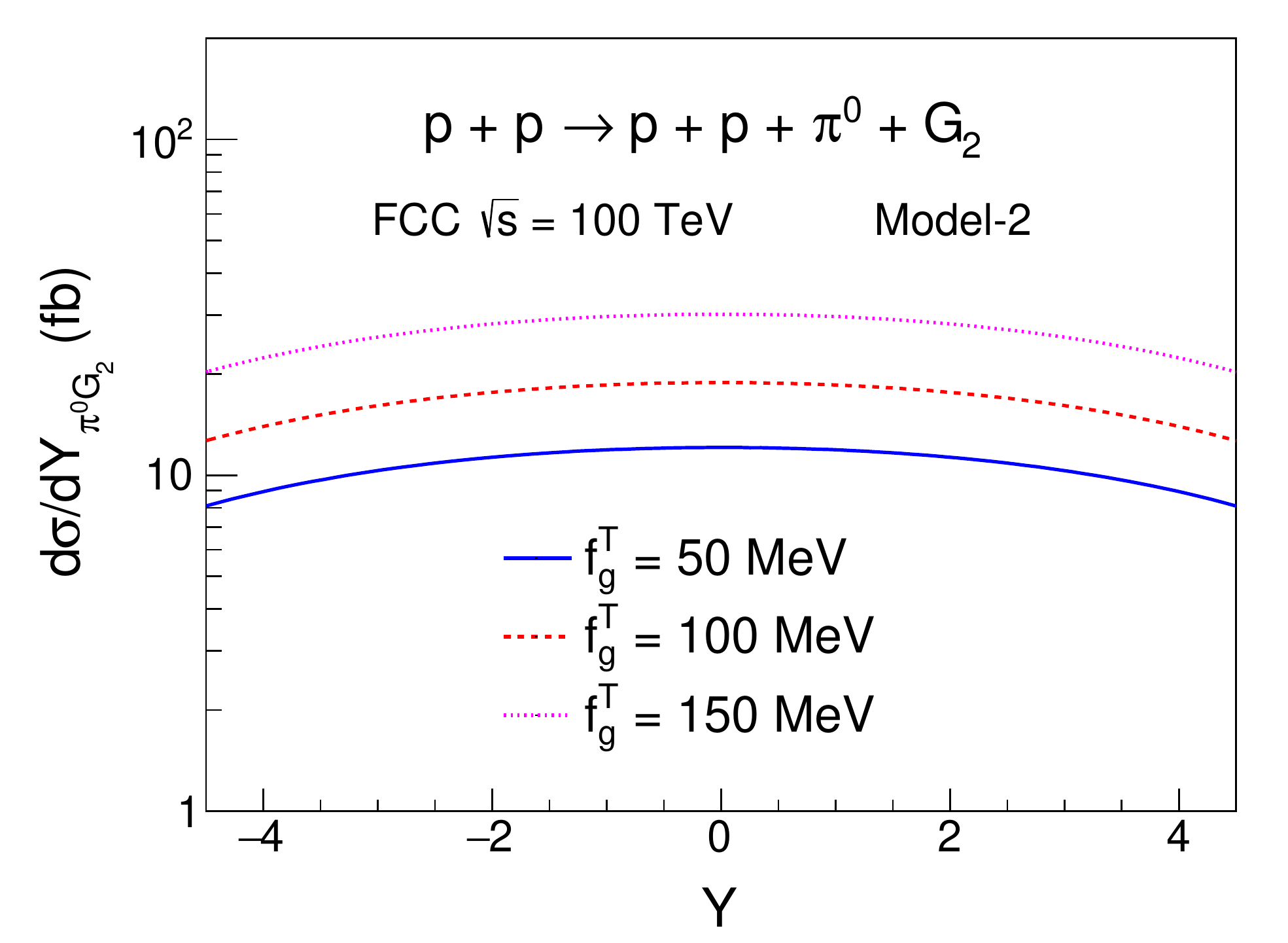}}
	\caption{Rapidity distributions for the UPC process $p+p\to p+p+G_2+\pi^0$ with model-1 (left panels) and model-2 (right panels). From top to bottom, it represents the predictions for the LHC, HE-LHC and FCC energies, respectively. The solid, dashed and dotted lines correspond to $f_g^T=50$ MeV, 100 MeV and 150 MeV \cite{Kivel:2017tgt}.}
	\label{fig:dsigmadY-p}
\end{figure}

\begin{figure}[H]
	\centering  
	\subfigure{
		\label{fig:pb1-mod1}
		\includegraphics[width=0.47\textwidth]{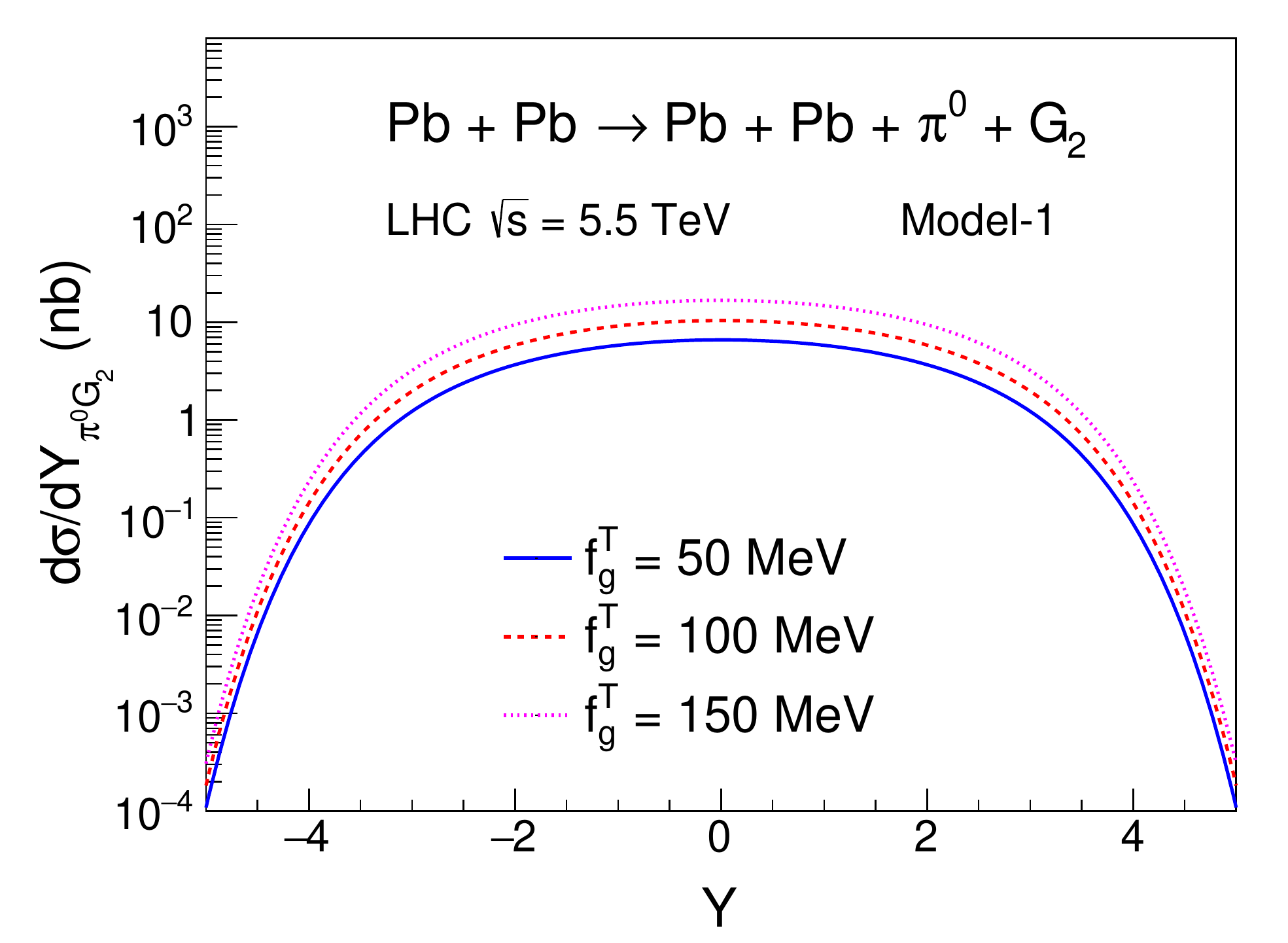}}
	\subfigure{
		\label{fig:pb1-mod2}
		\includegraphics[width=0.47\textwidth]{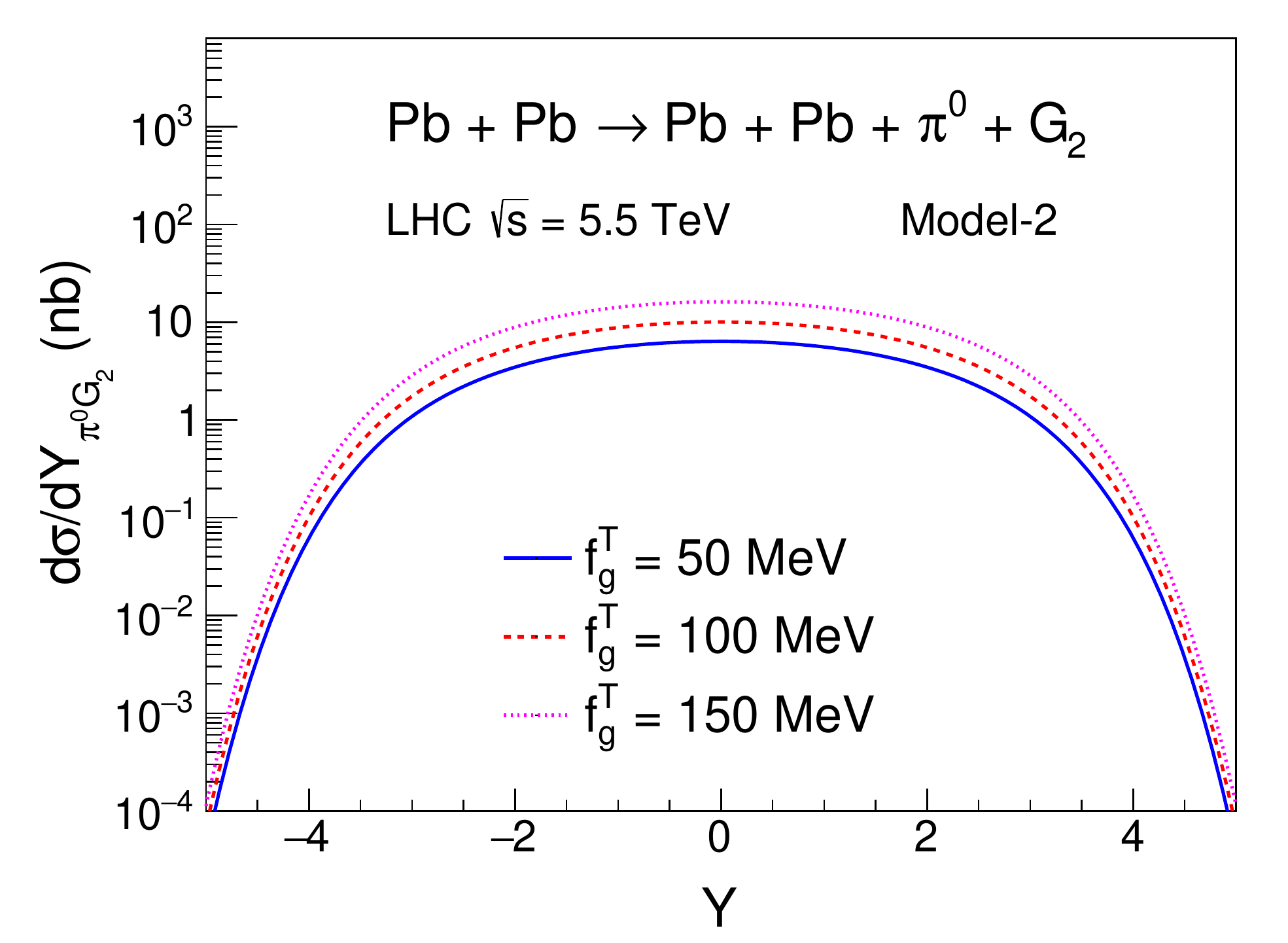}}
	\subfigure{
		\label{fig:pb2-mod1}
		\includegraphics[width=0.47\textwidth]{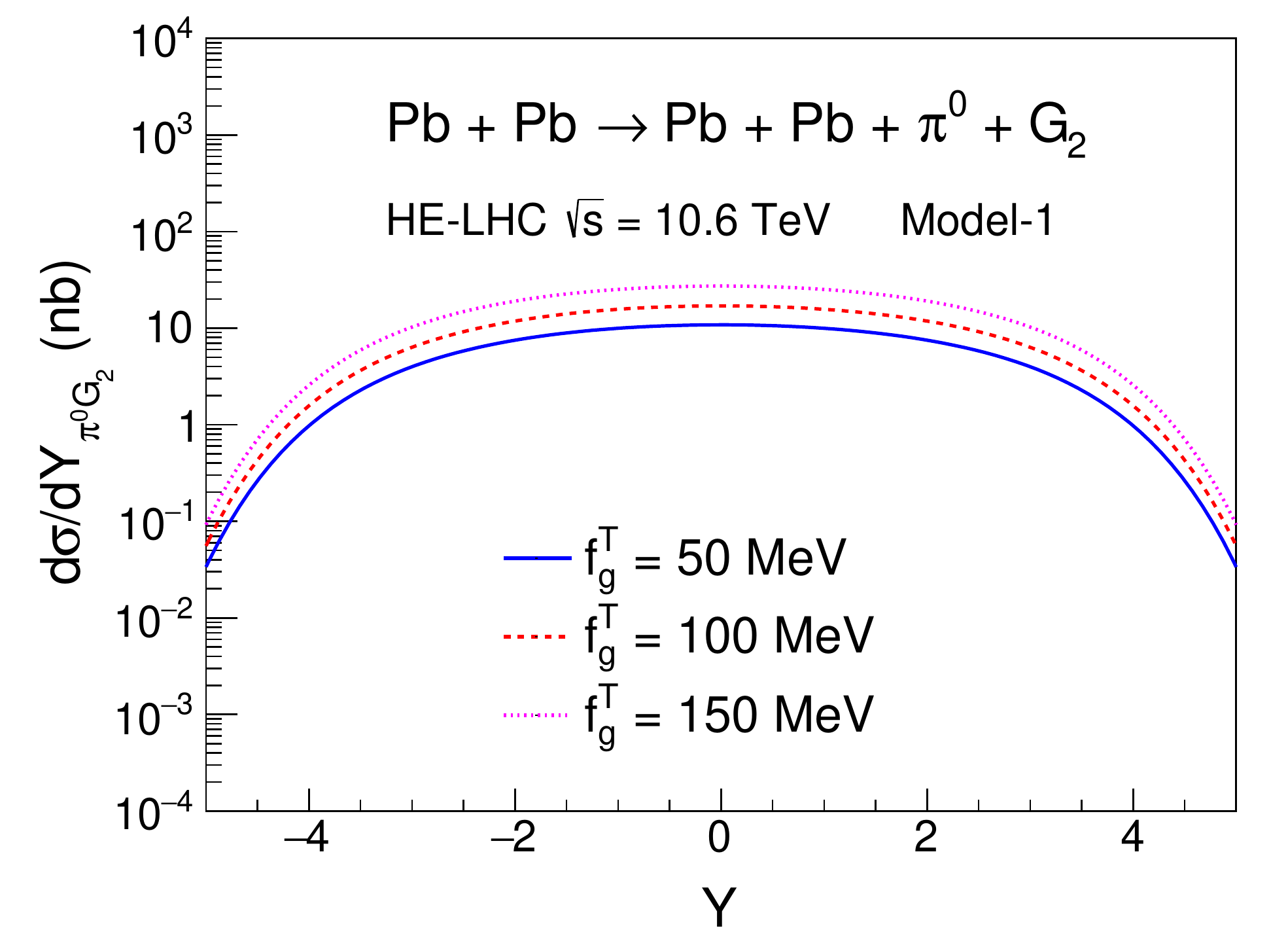}}
	\subfigure{
		\label{fig:pb2-mod2}
		\includegraphics[width=0.47\textwidth]{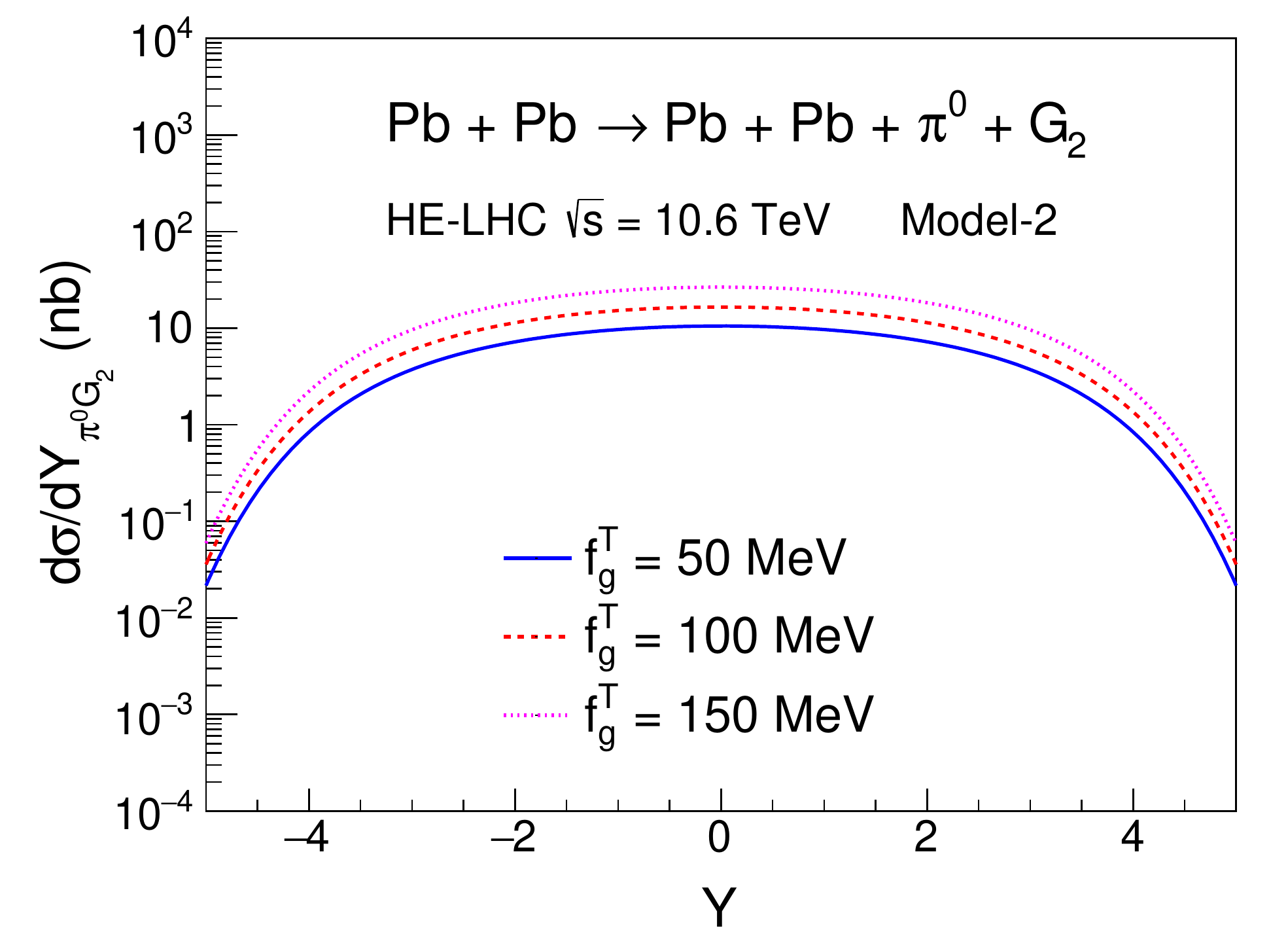}}
	\subfigure{
		\label{fig:pb3-mod1}
		\includegraphics[width=0.47\textwidth]{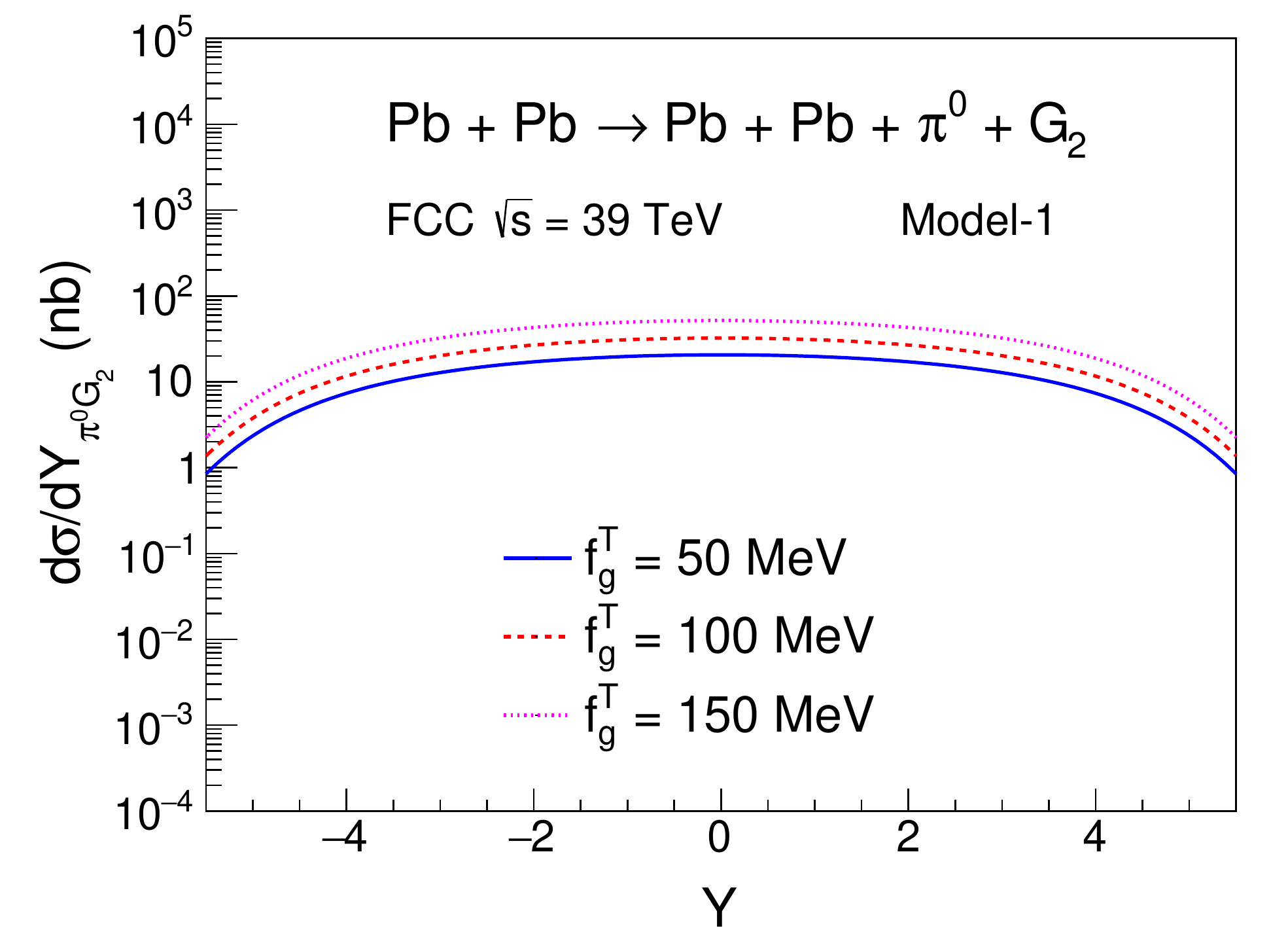}}
	\subfigure{
		\label{fig:pb3-mod2}
		\includegraphics[width=0.47\textwidth]{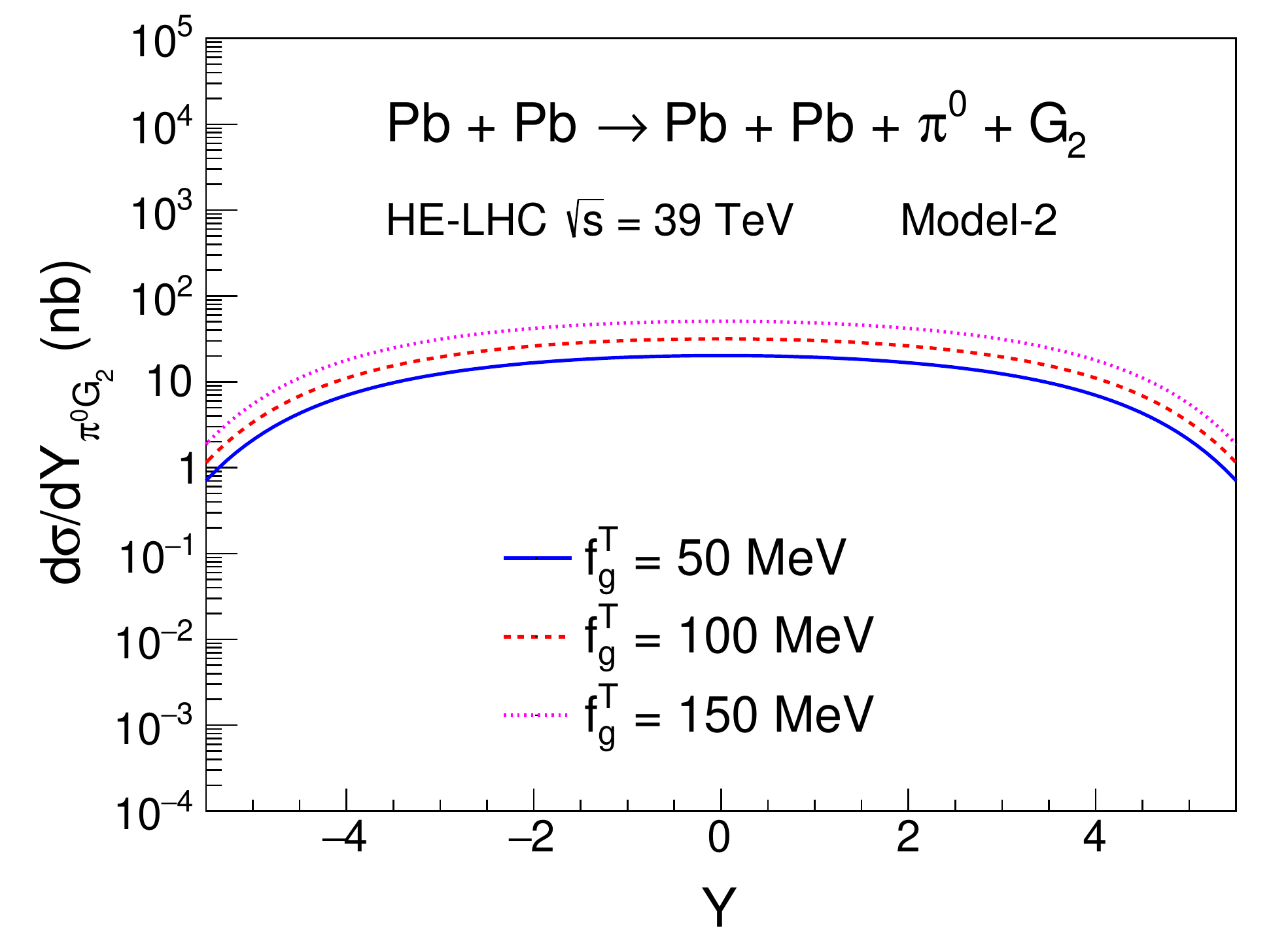}}
	\caption{Rapidity distributions for the UPC process $Pb+Pb\to Pb+Pb+G_2+\pi^0$ with model-1 (left panels) and model-2 (right panels). From top to bottom, it represents the predictions for the LHC, HE-LHC and FCC energies, respectively. The solid, dashed and dotted lines correspond to $f_g^T=50$ MeV, 100 MeV and 150 MeV \cite{Kivel:2017tgt}. }
	\label{fig:dsigmadY-pb}
\end{figure}

Next we give the results for UPCs. Considering the parameter selection of $\gamma\gamma\to G_2\pi^0$ cross section, we calculate the total cross section of UPCs in two ways, model-1 and model-2 mentioned in the previous section, respectively. We consider $AuAu$ collisions on RHIC and $pp$ and $PbPb$ collisions on LHC, respectively. We go further and consider the planned center-of-mass energies for the next run of the LHC, as well for the future High-energy LHC \cite{FCC:2018bvk} and Future Circular Collider \cite{FCC:2018vvp}. The
results presented in FIGs. \ref{fig:dsigmadY-au} to \ref{fig:dsigmadY-pb}. It is seen that there are only minor differences between model-1 and model-2. The selection of different coupling constants in each sub-figure can be significantly different and the trend is consistent with the FIG. \ref{fig:dsigma-2gamma}. It is because the differential cross section (\ref{eq:diff-xsec}) is proportional to the square of $f_g^T$. Moreover, we
have that for larger energies the rapidity distribution increases and becomes wider in rapidity. As expected from our previous discussions, the distribution is symmetric for $pp$ and $PbPb$ collisions. If one compares the FIGs. \ref{fig:dsigmadY-p} and \ref{fig:dsigmadY-pb}, it is clear that the value of predictions for midrapidities ($Y\simeq0$) increase with the energy and are a factor $\simeq10^6$ larger in $PbPb$ collisions than in $pp$ collisions. The enhancement in $PbPb$ collisions
is mainly associated to the $Z^4$ - factor in the nuclear photon flux \textsc{}\cite{Coelho:2020lyd}.

In the Tables \ref{tab:xsection-p} and \ref{tab:xsection-Pb} we present our predictions for the total cross sections and number of events, which includes the results of both model-1 and model-2 calculations (see Eqs. (\ref{eq:mod1}) and (\ref{eq:mod2})). Here we set the previously mentioned coupling constant $f_g^T=100$ MeV. Because of the low energy for RHIC, the total cross section is small and not shown in text. For $pp$ and $PbPb$ collisions, the expected integrated luminosities we use to estimate the number of events come from Refs. \cite{FCC:2018bvk,FCC:2018vvp} for the next planned run of LHC, HE-LHC and FCC. Moreover, we will also take
into account the tensor glueball decay mode, with the associated branching ratio for the processes $G_2\to\phi\phi$ being $1.74\%$ \cite{Yang:2013xba,BESIII:2016qzq}. The uncertainty of the branching ratio is not considered. We need to emphasize that there are fewer existing studies related to the branching ratio of $G_2\to\phi\phi$, thus we try to combine the lattice calculations with the experimental measurements presented by BESIII collaboration, which the tensor glueball is treated as tensor meson $f_2(2340)$. In principle, with fixed coupling $f_g^T$, both Tables \ref{tab:xsection-p} and \ref{tab:xsection-Pb} show the difference between model-1 and model-2. Although the difference is small, one could see that the calculation result of model-2 is smaller than that of model-1. 

Some comments are in order. First, in our analysis we only consider one of the possibilities of production of  tensor glueball from two-photon collisions, other photoproduction processes such as those mentioned above and the ``Pomeron-Pomeron exchange" are not considered in detail. These contributions cannot be ignored and serve as background processes for UPCs. For a complete discussion we suggest that readers refer to Refs. \cite{Schramm:1999tt,Machado:2010qg}. Second, we provide some theoretical uncertainties about the calculation of $\gamma\gamma\to G_2\pi^0$ matrix element and UPCs. In addition to the uncertainties analysis already mentioned, we need to emphasize other sources of theoretical uncertainties for double photon collisions \cite{Kivel:2017tgt}, which are provided by various subleading corrections. Besides, we also note the hadronic survival probabilities mentioned in Ref. \cite{Shao:2022cly}, which corresponds to the probability
that both scattered protons do not dissociate due to secondary soft hadronic interactions \cite{Dokshitzer:1987nc}. These factors will also modify the calculation of the UPCs cross section. Finally, we would like to declare something about the number of events. We emphasize that this work only gives results for the ideal case, the parameters and properties of the corresponding detectors are not the focus of this work. As shown in Tables \ref{tab:xsection-p} and \ref{tab:xsection-Pb}, the number of events produced by $pp$ collisions is generally greater than that of $PbPb$ collisions. This is due to the integrated luminosities of $pp$ beam, which is higher than that of heavy ions. However, there is likely no possibility to observe the low-$p_T$ decay products
of the very low-mass $G_2$ glueball (and $\pi^0$) decays in the ATLAS/CMS detectors as those
experiments are designed to trigger on and to reconstruct particles at much higher
transverse momenta/masses (typically above $p_T$, $m\sim5$ GeV). In addition, higher luminosity implies larger pileup that will only make impossible this measurement in $pp$ collisions (see the review article \cite{Soyez:2018opl}). Thus the absence of pileup in $PbPb$ collisions may make this system more beneficial to trigger on and reconstruct any UPCs compared to $pp$ collisions.

\section{Summary and outlook}
\label{sec:summary}
As a summary, in this work we have predicted the production of tensor glueball $G_2$ in $AuAu$, $pp$ and $PbPb$ collisions at UPCs. The two-photons interactions are used to construct the cross section of $\gamma\gamma\to \pi^0G_2$ process, thus we use it in UPCs ($h_1h_2 \rightarrow h_1 h_2 \pi^0G_2$). We also provide the  rapidity distributions and total cross sections in $pp$ and $PbPb$ collisions at UPCs and the corresponding number of events. Although the $pp$-collision case has higher integrated luminosities, the large pileup phenomenon previously mentioned makes it more difficult to reconstruct the final state tensor glueball. In contrast, $PbPb$ collisions have the opportunity to reconstruct the production of tensor glueball. From our estimation, one can expect that the $PbPb$ UPCs process could already produce about 10 signal events of $G_2\to \phi\phi$ when the integrated luminosity is $10$ nb$^{-1}$ at the LHC. Likely, the ALICE/LHCb experiments have better possibilities, to reconstruct the $G_2+\pi^0$ final state at low masses. Finally, we expect that future $PbPb$ UPC experiments performed by HE-LHC and FCC will provide higher statistics of tensor glueball production events.

\begin{acknowledgments}
	We are indebted to Ya-Ping Xie for discussions and comments. This work is supported by the Strategic Priority Research Program of Chinese Academy of Sciences under the Grant NO. XDB34030301 and
	Guangdong Major Project of Basic and Applied Basic
	Research NO. 2020B0301030008.
\end{acknowledgments}


\begin{table}[H]
	\caption{Cross sections and associated number of events in the tensor glueball decay mode $G_2\to \phi\phi$ for the tensor glueball production in $pp$ collisions. The integrated luminosities selection come from \cite{FCC:2018bvk,FCC:2018vvp}. The table includes the results of both model-1 and model-2 calculations. Here we set the previously mentioned coupling constant $f_g^T=100$ MeV.}
	\label{tab:xsection-p}
	\setlength{\tabcolsep}{2mm}{
	\begin{ruledtabular}
	\begin{tabular}{cccccc}
		&                       & \multicolumn{2}{c}{$\sigma_{tot}$ (fb)}     & \multicolumn{2}{c}{$\#$events ($G_2\to \phi\phi$)} \\ \hline
		$pp$ \textbf{collisions} & $\int\mathcal{L}$ & \textbf{model-1} & \textbf{model-2} & \textbf{model-1} & \textbf{model-2} \\ \hline
		LHC $\sqrt{s}=14$ TeV    & 3 ab$^{-1}/$20 years  & $8.48$ & $8.25$ & 442.20                   & 428.51                  \\
		HE-LHC $\sqrt{s}=27$ TeV & 15 ab$^{-1}/$20 years & $12.03$ & $11.72$ & 3127.08                  & 3059.66                 \\
		FCC $\sqrt{s}=100$ TeV   & 25 ab$^{-1}/$25 years & $20.57$ & $20.12$ & 8904.33                  & 8734.73                 \\ 
	\end{tabular}
	\end{ruledtabular}}
\end{table}

\begin{table}[H]
	\caption{Cross sections and associated number of events in the tensor glueball decay mode $G_2\to \phi\phi$ for the tensor glueball production in $PbPb$ collisions. The integrated luminosities selection $\int\mathcal{L}$ come from \cite{FCC:2018bvk,FCC:2018vvp}. The table includes the results of both model-1 and model-2 calculations. Here we set the previously mentioned coupling constant $f_g^T=100$ MeV.}
	\label{tab:xsection-Pb}
	\setlength{\tabcolsep}{2.5mm}{
		\begin{ruledtabular}
	\begin{tabular}{cccccc}
		\multicolumn{1}{l}{} & \multicolumn{1}{l}{} & \multicolumn{2}{c}{$\sigma_{tot}$ (nb)} & \multicolumn{2}{c}{$\#$events ($G_2\to \phi\phi$)} \\ \hline
		$PbPb$ \textbf{collisions}          & $\int\mathcal{L}$   & \textbf{model-1} & \textbf{model-2} & \textbf{model-1} & \textbf{model-2} \\ \hline
		LHC $\sqrt{s}=5.5$ TeV     & 10 nb$^{-1}/$experiment   & 44.50   & 42.46   & 7.8    & 7.4    \\
		HE-LHC $\sqrt{s}=10.6$ TeV & 10 nb$^{-1}/$month  & 88.23   & 84.35   & 15.79   & 14.63   \\
		FCC $\sqrt{s}=39$ TeV      & 110 nb$^{-1}/$month & 218.89  & 212.94  & 419.31  & 407.36  \\ 
	\end{tabular}
	\end{ruledtabular}}
\end{table}

\bibliographystyle{apsrev4-1}
\bibliography{refs}
\newpage

\end{document}